\documentclass[graybox]{svmult}

\usepackage{mathptmx}       
\usepackage{helvet}         
\usepackage{courier}        
\usepackage{type1cm}        
\usepackage{makeidx}         
\usepackage{graphicx}        
\usepackage{multicol}        
\usepackage[bottom]{footmisc}

\usepackage{natbib}
\newcommand{\apj}{ApJ}
\newcommand{\apjl}{ApJ}
\newcommand{\solphys}{Sol. Phys.}
\newcommand{\aap}{A \& A}

\newcommand{\rh}{\varrho}

\renewcommand{\vec}[1]{\mbox{\boldmath$#1$}}
\newcommand{\mean}[1]{\langle{#1}\rangle}
\newcommand{\fluc}[1]{{#1}^{\prime}}

\makeindex             

\begin{document}

\title*{Solar convection zone dynamics}
\author{Matthias Rempel}
\institute{Matthias Rempel\at HAO, National Center for Atmospheric Research, Boulder CO, USA \email{rempel@hao.ucar.edu}}
%
%
\maketitle

\abstract{
A comprehensive understanding of the solar magnetic cycle requires 
detailed modeling of the solar interior including the
maintenance and variation of large scale flows (differential rotation and
meridional flow), the solar dynamo and the flux emergence process
connecting the magnetic field in the solar convection zone with magnetic field
in the photosphere and above. Due to the vast range of time and length scales
encountered, a single model of the entire convection zone is still out of 
reach. However, a variety of aspects can be modeled
through a combined approach of 3D MHD models and simplified descriptions.
We will briefly review our current theoretical understanding of these 
processes based on numerical models of the solar interior.}

\section{Introduction}
\label{rempel-sec:1}
The solar convection zone comprises the outer most $30\,\%$ of the solar radius
and contains about $2\,\%$ of the total solar mass. Due to a density variation 
of more than 6 orders of magnitude a variety of different physical regimes are 
encountered. While fluid motions are are highly 
subsonic ($\mbox{Ma}\approx 10^{-4}$) 
and strongly influenced by rotation near the 
base of the convection zone, they turn supersonic in the photosphere and
the influence of rotation diminishes. The pressure scale height varies between
about 50 Mm at the base of the convection one and about 100 km in the 
photosphere of the sun. As a consequence a comprehensive model of the entire
convection zone is currently out of reach and different aspects 
have to be modeled independently. The deep convection
zone up to about 10-20 Mm beneath the photosphere can be modeled most 
efficiently using the anelastic approach which is filtering out sound waves, 
but is fully considering the compressibility in the stratification
\citep{Glatzmaier:1984}. The upper most parts of the convection zone require 
fully compressible MHD 
\citep[see e.g.][for a recent review]{Nordlund:etal:2009}. While most
anelastic models of the solar interior are global models with computational
domains covering an entire shell between two radii (or at least a shell 
segment), MHD models of the outer parts of the convection zone typically focus
on details in rectangular computational domains.

Apart from 3D MHD models adapted to the different physical regimes a
variety aspects have been modeled based on simplified models, such as the
mean field approach. Here the focus is on the large scales, while the effects
of unresolved turbulence is parametrized. Non-linear terms in the
momentum, energy and induction equations lead to non-vanishing second order 
correlation terms of small scale quantities that act as drivers for large
scale flows or as turbulent induction effects for the large scale magnetic 
field. The decomposition into large and small scale properties and the arising
correlation terms driving large scale flows are the strength and the weakness 
of this approach at the same time. On one hand the computational expense
is decreased by orders of magnitude allowing for simulations covering long 
time scales as well as exploring wide parameter ranges, on the other hand the
results are heavily dependent on parametrization of the second order
correlation terms. For a comprehensive description of mean field theory
we refer to \cite{Ruediger:Hollerbach:2004}. 
 
\section{Differential rotation and meridional flow}
\label{rempel-sec:2}
Differential rotation is the consequence of angular momentum transport in the
solar convection zone. Starting with a decomposition of the turbulent velocity 
field into fluctuating and (axisymmetric) mean flows 
$\vec{v}=\mean{\vec{v}}+\fluc{\vec{v}}$ leads to the following terms in the
angular momentum flux (neglecting magnetic terms for simplicity):
\begin{equation}
  \mean{F_i}=r\sin\theta\rh\,(
    \underbrace{\mean{\fluc{v_i}\,\fluc{v_{\phi}}}}_{\mbox{Reynolds stress}}
  +\underbrace{\mean{v_i}\Omega\,r\sin\theta}_{\mbox{Meridional flow}})\;.
  \label{rempel-eq1}
\end{equation}
Here the Reynolds stress describes the transport due to correlations of 
fluctuating turbulent velocity components, while the Meridional flow describes
the transport due to large scale coherent mean flows in the $r-\theta$ plane.
Angular momentum transport through Reynolds stresses requires the presence of
rotation and anisotropy and expressions for these 
transport terms have been derived within the mean field approach by 
\citet{Durney:Spruit:1979,Hathaway:1984} and more recently by  
\cite{Kitchatinov:Ruediger:1993} using a quasi-linear approach (see also
\cite{Kitchatinov:Ruediger:2005} for an improved representation).
In 3D simulations the influence of rotation on convection leads to a 
preferential north-south alignment of convection cells 
\citep{Gilman:1979,Miesch:etal:2000,Brun:Toomre:2002,Miesch:etal:2008}.
The consequence is a dominance of east-west motions over north-south
motions. By means of the Coriolis force eastward (faster rotating) flows are 
deflected equatorward, while westward (slower rotating) flows are deflected 
poleward, leading on average to an equatorward transport of angular momentum.

\begin{figure}
  \resizebox{\linewidth}{!}{\includegraphics{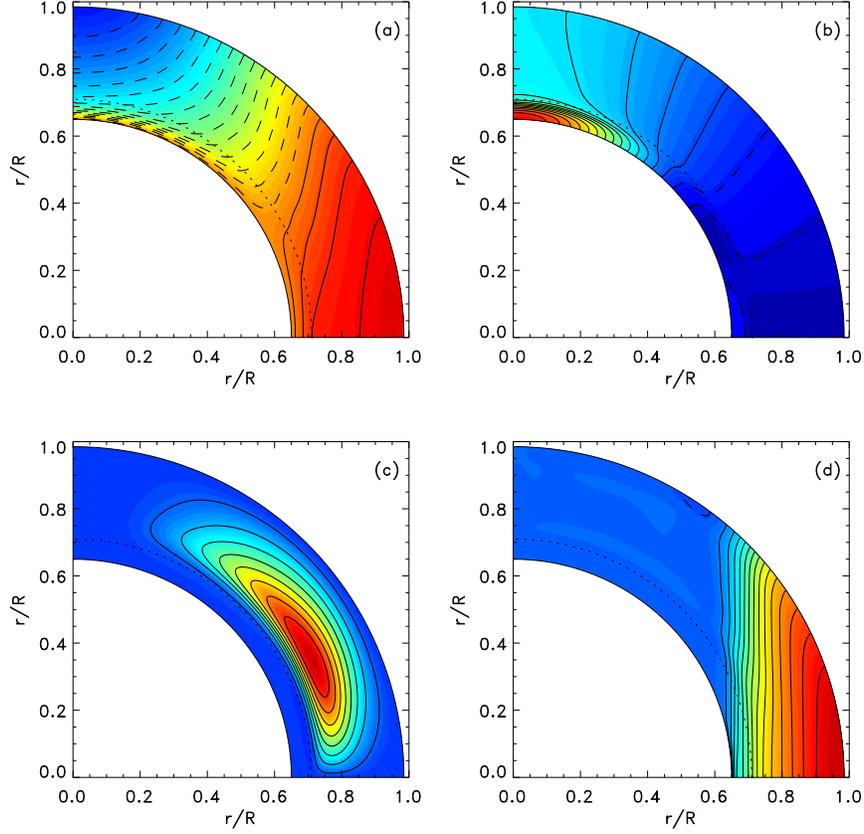}}
  \caption{Contour plots of differential rotation (a), entropy perturbation (b)
    and stream function of meridional flow (c) using the mean field model of
    \citet{Rempel:2005}. Panel (d) shows the differential
    rotation profile obtained using the same parametrization of the Reynolds 
    stress but neglecting the effects of baroclinicity.}
  \label{rempel-f1}
\end{figure}

The profile of differential rotation cannot be determined on the basis of
angular momentum transport processes alone. As stationary state requires beside
vanishing divergence of the total angular momentum flux also a force balance
in the meridional plane between Coriolis, centrifugal, buoyancy and pressure
forces. The latter is most conveniently expressed by (follows from 
$\phi$-component of vorticity equation):
\begin{equation} 
  r\sin\theta\frac{\partial\Omega^2}{\partial z}=
  \frac{g}{c_p\,r}\frac{\partial s}{\partial \theta}
  \label{rempel-eq2}
\end{equation}
Helioseismic observations by \cite{Thompson:etal:1996:science}
show clearly a differential rotation profile with contours of constant
$\Omega$ inclined by about $25\deg$ with respect to the rotation axis
(deviation from Taylor-Proudman state). It turns out that avoiding
the Taylor-Proudman state is a key problem for a theoretical understanding of
solar differential rotation. While early models attempted
to achieve this by assuming large viscosities (`Taylor-number puzzle' after 
\citet{Brandenburg:etal:1990:puzzle}), \citet{Kitchatinov:Ruediger:1995}
 showed that 
an alternative solution of this problem can be given if the anisotropic 
convective energy transport is  considered, leading to a pole-equator
temperature difference of about $10$ K. 
Anisotropic convective energy transport is automatically considered in global
3D simulations, but in many cases it turns out to be insufficient 
for obtaining solar-like differential rotation.

Recently \citet{Rempel:2005} showed that coupling between the tachocline
and convection zone can also provide the latitudinal entropy variation needed to
explain the observed profile of solar differential rotation. A typical 
solution from that model is shown in Fig.~(\ref{rempel-f1}), displaying
differential rotation (a), corresponding entropy perturbation (b) and the
stream function of the meridional flow (c). Panel (d) shows for comparison the
profile of differential rotation obtained if the effects of the entropy 
perturbation displayed in (b) are neglected. Inclusion of this effect through
the bottom boundary condition in 3D models allows also for more solar-like
differential rotation in 3D convection models \citep{Miesch:etal:2006}

While there is a general agreement that thermal effects are essential for
solar-like differential rotation, it is still unclear whether the required
latitudinal entropy variation is a consequence of anisotropic convective 
energy transport, imposed by the tachocline, or a combination of both.

How does the meridional flow come into play here? A stationary solution requires
that the divergence of the angular momentum flux Eq. (\ref{rempel-eq1})
vanishes. While in very special situations the Reynolds stress can be 
divergence free on its own, in general a contribution from the meridional
flow is required to close the system. It turns out that primarily the component
of the Reynolds stress that transports angular momentum parallel to the axis of
rotation influences most strongly the direction of the meridional flow. If
the transport of angular momentum is inward directed, the resulting meridional
flow is poleward at the surface and equatorward near the bottom of the 
convection zone. While this is found in most mean field models such as
\citet{Kitchatinov:Ruediger:1995}, 3D simulations present a more complicated 
situation. Early models with lower resolution \citep{Brun:Toomre:2002} 
typically show multi-cellular flow pattern, while a recent high resolution run
\citep{Miesch:etal:2008} shows a single flow cell (poleward at top, equatorward
at bottom of CZ) in the bulk of the convection zone. To which degree these 
results are converged with respect to numerical resolutions remains to be
seen in the future.

Differential rotation shows also cyclic variations known as torsional 
oscillations, which point toward a close relation to the solar magnetic cycle. 
We refer here to \citet{Howe:2009} and \citet{Brun:Rempel:2009} for 
reviews of observations as well as theoretical models for the time varying 
component of solar differential rotation.   

\section{Solar Dynamo}
\label{rempel-sec:3}
Similar to models of differential rotation and meridional flow we
discussed in section \ref{rempel-sec:2}, also the solar dynamo is modeled
through a combination of mean field models and 3D simulations. Currently
mean field models of the solar dynamo are the only models that provide dynamo 
solutions that are compatible with basic cycle features and can be evolved
over time scales much longer than a cycle. However, as already stated above,
these models are heavily dependent on parametrization of turbulent induction
processes and cannot provide an explanation from first principles. 
On the other hand, 3D MHD simulations describe 
currently only aspects of the dynamo process, a comprehensive model of a 
solar dynamo with features compatible with the basic dynamo constraints 
derived from the solar butterfly diagram is still an open challenge.  

Regardless of the adopted modeling approach the primary uncertainties regarding
the underlying dynamo process are similar. Many of these uncertainties result
directly from our limited ability to model processes from first principles
and the rather sparse observational constraints on the solar interior. The
best known ingredient is differential rotation ($\Omega$-effect) due to 
observational constraints
from helioseismology on the mean profile and variation of differential rotation
\citep{Howe:2009}. But even the exact knowledge of the differential rotation
profile is not sufficient to determine whether radial velocity gradients at
the base of the convection zone (tachocline) or latitudinal gradients in the
bulk of the convection zone play the major role in the generation of toroidal
magnetic fields, since this would require knowledge of the detailed 
distribution of poloidal field in the convection zone. Even less known are the
processes related to the regeneration of poloidal field ($\alpha$-effect).
A third unknown are the transport processes of magnetic flux in the convection
zone. Since in general the locations where the $\alpha$-effect and 
$\Omega$-effect 
operate do not coincide, transport of magnetic flux inbetween these regions
is crucial for a coherent operation of the large dynamo in the convection zone.

It is beyond the scope of this paper to review all the possible dynamo 
scenarios which have been considered and we refer to  \citet{Charbonneau:2005}
for further reading. In the following three subsections we will briefly 
discuss some of the key uncertainties and open questions.

\subsection{Role of tachocline}
Soon after helioseismology revealed the detailed structure of differential
rotation in the solar interior \citep{Thompson:etal:1996:science}
it was suggested that the base of the 
convection zone with its strong radial shear layer (tachocline) is a likely
location for the solar dynamo (production of strong toroidal field by
shear). In addition, the stable stratification found in the solar 
overshoot region at the base of the convection zone allows for storage of 
magnetic field over time scales comparable 
to the solar cycle. Both aspects are crucial since simulations of 
rising flux in the convection zone as well as studies of magnetic 
stability in the solar overshoot regions (see section \ref{rempel-sec:4} for 
further detail) point toward a rather strong toroidal magnetic field of 
$10^5$ Gauss at the base of the convection zone.
More recently the role of the tachocline for the organization and 
amplification of large scale toroidal field has been also seen in global
3D MHD simulations of the solar dynamo \citep{Browning:etal:2006}. However,
\citet{Brown:etal:2009} presented simulations of solar like stars at faster 
rotation rates, which point toward the possibility that substantial magnetic 
field can be produced and maintained within the convection zone in near 
equatorial regions. It is currently not clear to which degree this result
can be also relevant for the solar rotation rate.

While most models of flux emergence point toward a field
strength of $10^5$ Gauss at the base of the convection zone, it is far from
trivial to amplify field to this strength solely through differential rotation.
Dynamo models that include non-linear feed-backs consistently 
\citep{Rempel:2006:dynamo} lead to an upper limit more around $10^4$ Gauss,
similar values are also found in most 3D simulations such as 
\citep{Browning:etal:2006}. Whether this discrepancy can be bridged through
an alternative field amplification mechanism (e.g. harvesting potential 
energy of the stratification as proposed by  \citet{Rempel:Schuessler:2001}) 
or the possibility that also initially weaker magnetic field from the bulk of 
the convection zone can lead to active region formation is currently an open
question.   

\subsection{Regeneration process of poloidal field}
The details of the 
processes rebuilding the poloidal field from toroidal field are still
very uncertain. In the meanfield language these processes are formally
described as $\alpha$-effect and in the context of solar dynamo models the 
following 3 classes of $\alpha$-effects are typically considered: 
1. Helical turbulence, 2. MHD shear flow instabilities in the tachocline,
3. Rising flux tubes in the convection zone (Babcock-Leighton). While all
these processes are likely to contribute, their amplitude and spacial 
distribution is not known well enough to clearly quantify their individual 
role. 

Furthermore recent research also points toward highly non-linear and
also time dependent $\alpha$-effects resulting from additional constraints 
due to conservation of magnetic helicity \citep{Brandenburg:Sandin:2004}.
Indirect constraints on the operation of the $\alpha$-effect might be gained
from helicity fluxes observable in the photosphere and above.

The only $\alpha$-effect contribution that is directly constrainable through 
observations is the 
Babcock-Leighton $\alpha$-effect \citep{Babcock:1961,Leighton:1969}, which 
has been used in most of the recent flux-transport dynamo models 
\citep{Dikpati:Charbonneau:1999,Dikpati:etal:2004,Rempel:2006:dynamo}. The
Babcock-Leighton $\alpha$-effect is based on the flux emergence process leading
to the formation of active regions, the key ingredient is the systematic tilt
resulting from the action of the Coriolis force twisting the rising flux 
tube. While it is possible to construct dynamo models entirely
based on the Babcock-Leighton $\alpha$-effect, these models lead in general to
rather strong polar fields at the surface in contradiction with observations,
unless a strong magnetic diffusivity gradient and additional contributions 
from $\alpha$-effects at the base of the convection zone are considered 
\citep{Dikpati:etal:2004}. 

\subsection{Transport of magnetic flux in convection zone}
Traditionally most models considered only turbulent transport in the 
convection zone, which can be decomposed (in the meanfield language) into 
diffusive transport (turbulent diffusion) but also advection like transport 
in form of turbulent pumping. The latter has been also studied extensively 
through  3D MHD simulations \citep{Tobias:etal:1998,Tobias:etal:2001}. 
If magnetic
field becomes sufficiently strong magnetic buoyancy drives additional
transport in terms of rising flux bundles that can lead to the formation of 
active regions on the visible surface (see section \ref{rempel-sec:4} for
more detail). Additional to these processes magnetic flux can be transported
by the large scale meridional flow in the convection zone. Dynamo models based
primarily on the latter are called flux transport dynamos and were first 
introduced by \citet{Choudhuri:etal:1995} and \citet{Dikpati:Charbonneau:1999} 
and have been developed further by several groups 
since then. 

The attraction of flux transport dynamos comes
primarily from the fact that a meridional flow which is poleward at the
top and equatorward at the bottom of the convection zone gives a very robust
explanation for the equatorward propagation of the activity in the course 
of the solar cycle. In addition the poleward flow in the near surface levels 
in combination with the systematic tilt angle of sunspot groups leads 
automatically to the correct phase relation between toroidal and poloidal 
field. However, as pointed out by \citet{Schuessler:2005:phase}, the phase 
relation is primarily a consequence of the tilt angle of active regions and 
in that sense only a weak constraint on dynamo processes in the
solar interior. For meridional flow speeds consistent 
with surface observations and an extrapolation based on mass conservation for 
the deeper layers, these models also yield dynamo periods in agreement with 
the solar cycle. 

Overall the flux transport picture is currently one of the most successful
scenarios for the large scale solar dynamo, but (as many other models) it
is based on two strong 
assumptions which cannot be proven from first principles: 
1. The meridional flow is dominated by one flow cell with 
poleward flow close to the surface layers and equatorward flow at the
base of the convection zone. 2. Turbulent transport processes are sufficiently
weak to allow advection effects to dominate. While meanfield models of
differential rotation and meridional flow typically lead to the required flow
patterns \citep[see e.g.][]{Kueker:Ruediger:2005:AN}, the situation is more 
complicated in 3D simulations. Most of the earlier models at moderate 
resolution lead to multi-cellular flows \citep{Brun:Toomre:2002}, in contrast
a more recent model at higher resolution shows a flow 
pattern dominated by a single cell in the convection \citep{Miesch:etal:2008}. 
Overall the situation cannot be considered as converged yet. The amplitude of 
turbulent transport estimated from simple mixing length arguments is typically 
1 - 2 orders of magnitude larger than the values required for the flux 
transport picture. Since turbulent transport is in general more complicated 
than a simple diffusive transport this aspect needs to be studied in more 
detail through 3D simulations taking into account the presence of large scale
flows and the full non-linearity of the problem .

\section{Flux emergence process}
\label{rempel-sec:4}
It is generally accepted that sunspots form from magnetic field rising from 
the base of the convection zone to the surface 
\citep[see reviews by][and further references therein]{Moreno-Insertis:1997b,
Fisher:etal:2000,Fan:2004},
Solar dynamo models as presented in section \ref{rempel-sec:3} focus on the 
large scale evolution of magnetic field an cannot address detailed processes
such as the flux emergence process leading to the formation of sunspots
on the visible surface of the sun. The latter is primarily a consequence of 
limited numerical resolution. Nevertheless, studying flux emergence is integral
to our understanding of solar magnetism, since it allows us to connect the solar
dynamo to observational constraints on the magnetic field structure in the
solar photosphere. To date the flux emergence process has been studied decoupled
from dynamo models using a variety of different approaches. 

\subsection{Flux emergence in lower convection zone}
Early work was based
on the thin flux tube approximation \citep{Choudhuri:Gilman:1987,Fan:etal:1993,
Fan:etal:1994,Moreno-Insertis:etal:1994,Schuessler:etal:1994,
Caligari:etal:1995}.
These studies concluded that the overall properties of active regions, such as
the low latitude of emergence, latitudinal trend in tilt angles as well as
asymmetries between leading and following spots can be understood on the basis
of rising flux tubes, provided the initial field strength at the base of the 
convection zone is around 100 kG. This conclusion was also consistent with
stability considerations of flux in a subadiabatic overshoot region 
\citep{FerrizMas:Schuessler:1993,FerrizMas:Schuessler:1995}. 

Based on two-dimensional MHD simulations it was early realized by 
\cite{Schuessler:1979:tubes} that untwisted magnetic flux tubes cannot 
rise coherently and fragment. It was shown later by 
\cite{Moreno-Insertis:Emonet:1996,Emonet:Moreno-Insertis:1998} that this 
fragmentation can be alleviated provided that flux tubes have enough initial
twist. 

More recently also 3D MHD simulations of rising flux tubes based on
the anelastic approximation have become possible \citep{Fan:2008} and give
support for results from earlier simulations based on the thin flux tube 
approximation. It was however found by \citet{Fan:2008} that there is a very
delicate balance between the amount of twist required for a coherent rise and
the amount of twist allowed to be in agreement with observations of sunspot
tilt angles (twist with the observed sign produces a tilt opposite to the 
effect of Coriolis forces on rising tubes).

The simulations presented above consider the flux emergence process decoupled 
from convection. First attempts to address
flux emergence in global simulations of the convection zone were made recently 
by \citet{Jouve:Brun:2007,Jouve:Brun:2009}. Understanding the interaction of
emerging flux with the ambient convective motions in the convective envelope
is a crucial step toward more realism; however, currently the focus on the 
global scale limits the resolution required to resolve this interaction in 
detail. Substantial progress will likely happen in the next decade with 
increase in computing power.
  
\subsection{Flux emergence in upper convection zone}
Another challenge is understanding the last stages of the flux emergence process
in the near surface layers (last 10 - 20 Mm). All of the models presented above 
exclude the upper most 10 - 20 Mm since the adopted approximations
(thin flux tube, anelastic) loose their validity and also the steep decrease
of pressure scale height and increase in convective velocities would lead
to very stringent resolution and time step constraints. The upper most
layers of the convection zone require fully compressible MHD and also 
the inclusion of radiative processes is necessary if a detailed comparison with
the available high resolution observations is desired
\citep{Cheung:etal:2007,Cheung:etal:2008,Cheung:etal:2010}. 
While the primary modeling focus
in the deep convection zone lies on large scale properties of
active regions, near surface simulations focus on the detailed interaction
of emerging flux with convection on the scale of granulation and below. One
of the major open questions concerns the re-amplification process of magnetic
flux into coherent sunspots from flux that has risen through a convection zone
with a density stratification of six orders of magnitude.

Recently MHD simulations with radiative transfer also provided a breakthrough
in our understanding of sunspot fine structure in the photosphere such as umbral
dots, penumbral filaments, light bridges and the Evershed flow in terms of a
common magneto-convection process modulated by inclination angle and field 
strength \citep{Schuessler:Voegler:2006,Heinemann:etal:2007,Rempel:etal:2009,
Rempel:etal:Science}.   

\subsection{Open questions, connection to dynamo models}
While we have seen strong progress in modeling the flux emergence process over
the past decades, we do not have at this point a fully consistent model. The
latter is a consequence of the fact that different aspects are modeled 
independently due to computational constraints. As a consequence there are
some 'missing links' between different models, which have to be addressed in
the future through a more coherent coupling of models. Here we mention just
a few of the open questions: 1. As pointed out before most models of emerging 
magnetic flux require an initial field strength of about $10^5$ Gauss at the
base of the convection zone to be consistent with observational constraints.
On the other hand the majority of dynamo models falls short of such values,
more typical are $10^4$ Gauss. 2. Due to the strong density stratification 
in the convection zone even magnetic field with initially $10^5$ is weakened
to sub kG field strength in the upper most layers of the convection during
the emergence process. It is currently an open question if such weak field
can get re-amplified to sunspot field strength. Near surface simulations
start very often from $10$ kG field about 5 Mm beneath the photosphere
\citep{Cheung:etal:2008} to overcome the influence of convective motions.
3. Rising $\Omega$-shaped flux tubes in the deep convection zone form 
typically as
low wave number instability ($m=1$ and $m=2$ modes are preferred). In the
near surface layers such low wave numbers should lead to much faster diverging
motions in bipolar sunspot groups as observed (due to magnetic tension) if 
sunspots stay connected to their deep roots. A possible dynamical 
'disconnection' mechanism has been proposed by \citet{Schuessler:Rempel:2005},
but it is also unclear if sunspots are sufficiently stable if they are rather
shallow. 

\section{Summary} 
We presented here a brief summary of the state of the art of modeling of
dynamical processes in the solar convection zone with focus on differential
rotation/meridional flow, the large scale solar dynamo and the flux emergence
process. We see currently in this field a dramatic change from more simplified
models toward large 3D MHD simulations, primarily driven by the strong
increase in computing resources. At the same time the field suffers 
from very limited observational constraints on processes in the solar
interior. Progress in the future will heavily rely on improving and
exploiting helioseismic constraints and also on coupling models to allow for
a check of consistency. In the near terms the latter is likely to be most 
successful for models of the flux emergence process. 

\begin{acknowledgement}
The National Center for Atmospheric Research is sponsored by the National
Science Foundation. This research  was partially funded through NASA award 
number NNH09AK14I. 
\end{acknowledgement}

\end{document}